\documentclass[sigconf,nonacm]{acmart}

\usepackage{enumitem}

\usepackage[linesnumbered,boxed,ruled,commentsnumbered]{algorithm2e}
\usepackage{xcolor}
\usepackage{multirow}
\usepackage{color}
\usepackage{colortbl}
\usepackage{balance} 
\colorlet{lightshadingone}{green!5}
\colorlet{lightshadingtwo}{blue!5}

\begin{document}

\title{Dynamic Network-Based Two-Stage Time Series Forecasting for Affiliate Marketing}

\author{Zhe Wang}
\affiliation{
  \institution{Xidian University}
  \department{School of Computer Science and Technology}
  \city{Xi'an}\country{China}
  }
\email{zwang_01@stu.xidian.edu.cn}

  \author{Yaming Yang}
\affiliation{
  \institution{Xidian University}
    \department{School of Computer Science and Technology}
  \city{Xi'an}\country{China}
  }
\email{yym@xidian.edu.cn}

\author{Ziyu Guan}
\affiliation{
  \institution{Xidian University}
    \department{School of Computer Science and Technology}
  \city{Xi'an}\country{China}
  }
\email{zyguan@xidian.edu.cn}

\author{Bin Tong}
\affiliation{
  \institution{Alibaba Group}
  \city{Hangzhou}\country{China}
  }
\email{tongbin.tb@taobao.com}

\author{Rui Wang}
\affiliation{
  \institution{Alibaba Group}
  \city{Hangzhou}\country{China}
  }
\email{qianmian.wr@taobao.com}

  \author{Wei Zhao}
  \authornote{Corresponding authors}
\affiliation{
  \institution{Xidian University}
    \department{School of Computer Science and Technology}
  \city{Xi'an}\country{China}
  }
\email{ywzhao@mail.xidian.edu.cn}	

\author{Hongbo Deng}
\authornotemark[1]
\affiliation{
  \institution{Alibaba Group}
  \city{Hangzhou}\country{China}
  }
\email{hbdeng@acm.org}

\renewcommand{\shortauthors}{Zhe Wang et al.}

\begin{abstract}
In recent years, affiliate marketing has emerged as a revenue-sharing strategy where merchants collaborate with promoters to promote their products. It not only increases product exposure but also allows promoters to earn a commission. This paper addresses the pivotal yet under-explored challenge in affiliate marketing: accurately assessing and predicting the contributions of promoters in product promotion. We design a novel metric for evaluating the indirect contributions of the promoter, called propagation scale. Unfortunately, existing time series forecasting techniques fail to deliver accurate predictions due to the propagation scale being influenced by multiple factors and the inherent complexities arising from dynamic scenarios. To address this issue, we decouple the network structure from the node signals and propose a two-stage solution: initially, the basic self-sales and network structure prediction are conducted separately, followed by the synthesis of the propagation scale. Specifically, we design a graph convolution encoding scheme based on descendant neighbors and incorporate hypergraph convolution to efficiently capture complex promotional dynamics. Additionally, three auxiliary tasks are employed: self-sales prediction for base estimations, descendant prediction to synthesize propagation scale, and promoter activation prediction to mitigate high volatility issues. Extensive offline experiments on large-scale industrial datasets validate the superiority of our method. We further deploy our model on Alimama platform with over $100,000$ promoters, achieving a $9.29\%$ improvement in GMV and a $5.89\%$ increase in sales volume.
\end{abstract}

% \begin{CCSXML}
% <ccs2012>
%    <concept>
%        <concept_id>10010405.10003550</concept_id>
%        <concept_desc>Applied computing~Electronic commerce</concept_desc>
%        <concept_significance>500</concept_significance>
%        </concept>
%    <concept>
%        <concept_id>10002951.10003317.10003347.10003350</concept_id>
%        <concept_desc>Information systems~Recommender systems</concept_desc>
%        <concept_significance>500</concept_significance>
%        </concept>
%  </ccs2012>
% \end{CCSXML}

% \ccsdesc[500]{Applied computing~Electronic commerce}
% \ccsdesc[500]{Information systems~Recommender systems}


\maketitle

\section{Introduction}

With the widespread adoption of the Internet and the rapid development of information technology, an increasing number of merchants have turned to online platforms. While online merchants possess a large number of products, they often lack effective sale channels to directly reach potential customers. In contrast, social networks have penetrated into people's daily lives, serving as a primary medium for information propagation, but the means of monetization remain limited. In response to these challenges, affiliate marketing is emerging~\cite{af-1,af-2,af-3}. As a Cost Per Sales (CPS) promotion model, affiliate marketing enables merchants to take advantage of the power of social networks to spread the product to a wider audience in a short period, while providing affiliate marketers (referred to as promoters) with commissions based on the quantifiable promotion effect (i.e, sales).

\begin{figure}[ht]
\centering
\includegraphics[width=1\columnwidth]{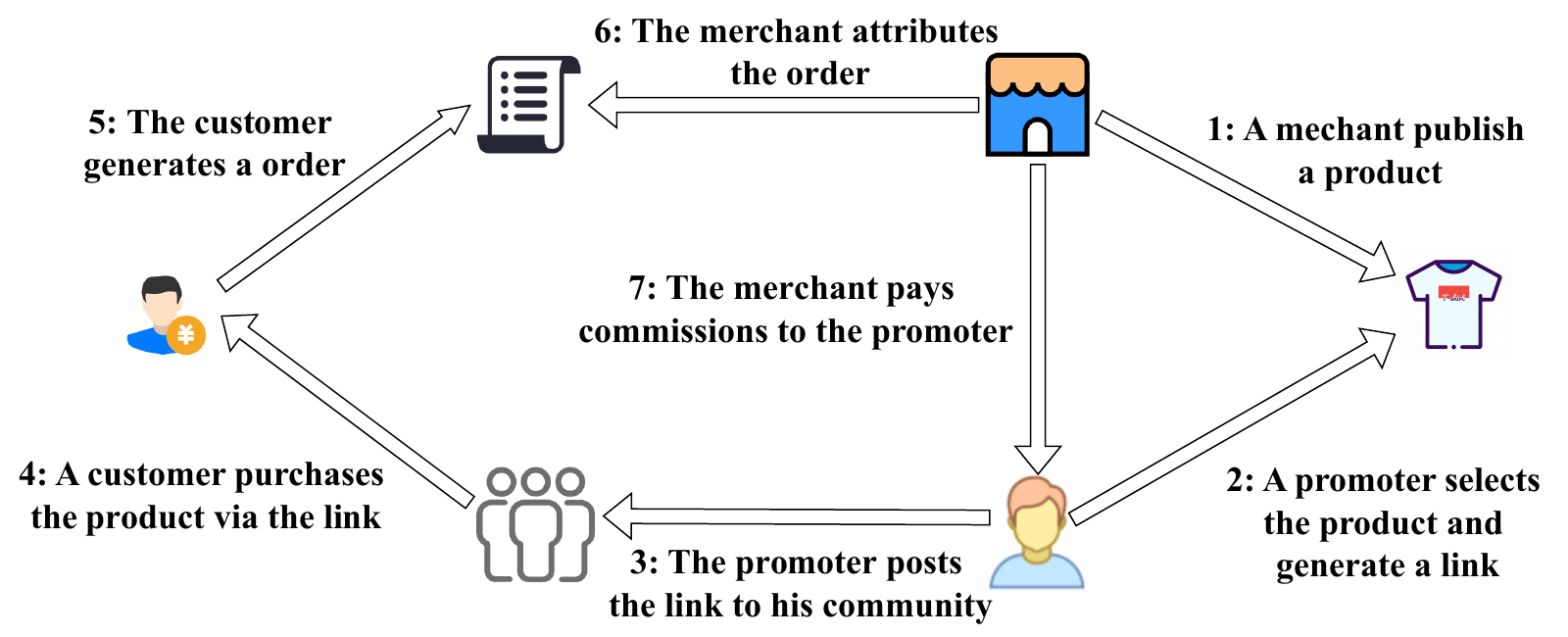}
\caption{A typical example for affiliate marketing.} 
\label{fig:case-1}
\end{figure}

Figure~\ref{fig:case-1} intuitively illustrates how affiliate marketing works by showing the interactions among several key roles. Initially, $\textit{merchants}$ publish their \textit{products}. Subsequently, $\textit{promoters}$ select products and generate links to share within their $\textit{communities}$. When a $\textit{customer}$ eventually makes a purchase through one of these product links, an $\textit{order}$ is created. The merchant then attributes the order to the promoter who generated the link and pays the appropriate commission. Affiliate marketing not only broadens the scope of product propagation and enhances the merchant's revenue, but also provides additional income for promoters, establishing a mutually beneficial online e-commerce ecosystem.

\begin{figure*}[ht]
\centering
\includegraphics[width=2.1\columnwidth]{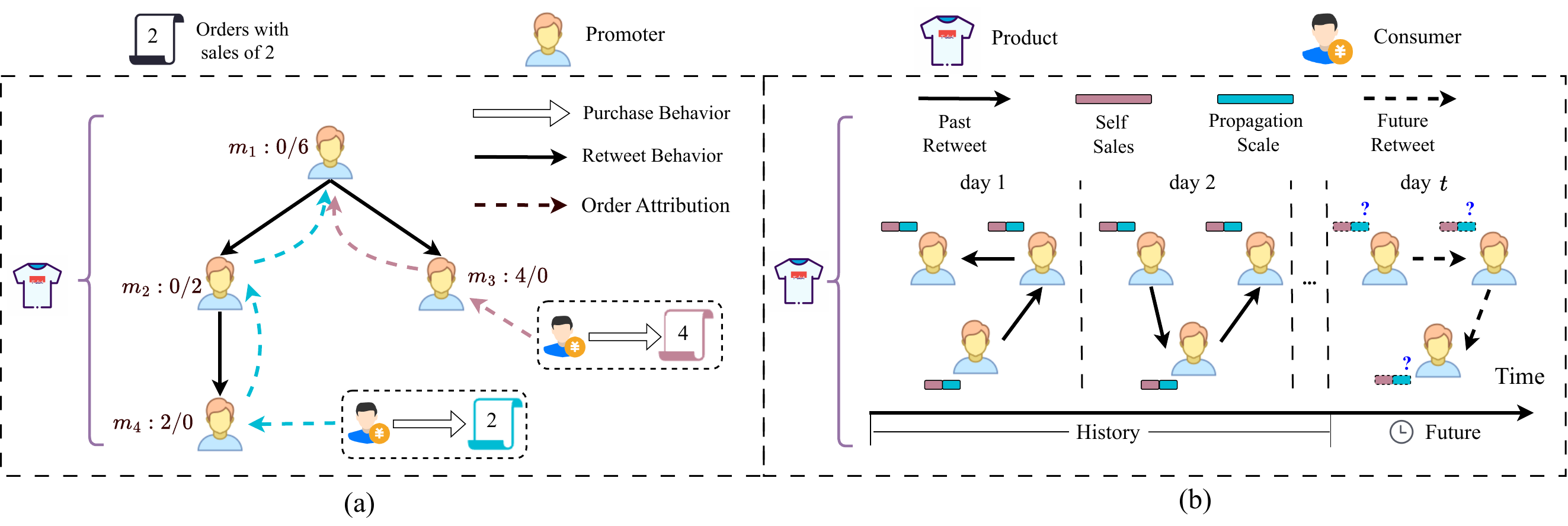}
\caption{ (a) A toy promotion network with self-sales (left) and propagation scale (right) separated by "/"; (b) Schematic of propagation scale prediction task, where the dynamically evolving promotion network is discretized by days.}
\label{fig:case-2}
\end{figure*}

% To enhance the operational efficiency of affiliate marketing, it is imperative to quantify the contributions of promoters in the propagation of specific products.
Promoters play a crucial role in affiliate marketing, serving as a bridge between merchants and consumers. Quantifying the promotional ability of promoters for specific products aids merchants in evaluating and effectively allocating promotional channel resources while addressing fairness issues in commission distribution. The current approach is to use the quantity of sales directly driven by promoters as the measurement standard (referred to as \textit{self-sales} in this paper), which is used to evaluate the direct promotion capability of promoters. However, it has a notable limitation, as it overlooks the indirect contributions of promoters. For instance, as shown in Figure~\ref{fig:case-2}(a), when customers purchase the product via the link generated by $m_4$, which is propagated from $m_2$, an order is created. Although $m_2$ has a self-sales of 0, it plays a crucial role in the promotion of the product, as it indirectly contributes to the generation of this order. To comprehensively evaluate a promoter's capabilities, it is essential to develop an additional metric to assess promoters' indirect promotion capability. In this paper, we define it as \textit{propagation scale}, which is equal to the sum of the sales corresponding to orders indirectly driven by the promoter. 

In this work, we focus on the more challenging task of predicting the propagation scale, as shown in Figure~\ref{fig:case-2}(b), which is essentially a time series forecasting task. However, existing techniques are inadequate for this task, as our experimental observations reveal that propagation scales fail to exhibit clear regularity, making direct prediction difficult. Meanwhile, existing GNN-based time series forecasting methods~\cite{dcrnn,stgcn,lsgcn,mtgnn,tpgnn} apply graph convolution directly to signals on static networks, making them incapable of handling dynamically evolving promotional networks.

We note that the propagation scale of a promoter can be derived from the self-sales of its descendants, where self-sales is an objective metric with relatively fewer influencing factors. For example, in Figure~\ref{fig:case-2}(a), the propagation scale of $m_2$ originates from the self-sales of $m_4$. This observation inspires us to propose a two-stage forecasting scheme: first, predict the fundamental self-sales and promotion network separately, and then synthesize the propagation scale based on them. In this way, graph information is independently modeled to enable adaptation to dynamic scenarios. To achieve this goal, we need to address two challenges.

% large network scale in quantity.
\textbf{Challenge 1:} Promotion networks dynamically evolve over time, and different products correspond to distinct networks during various periods, resulting in numerous different networks. This makes predicting all network architectures impractical.

\textbf{Challenge 2:} Since the target audience of each product is limited, the number of promoters participating in specific product propagation is significantly smaller than that of all promoters. Additionally, the promotion scale of a product is susceptible to business strategies. For instance, price reductions or increased commission may trigger a surge in its promotion.

To address Challenge $1$, we modify our task to eliminate the need for predicting network details but instead focus on identifying descendant sets. At the same time, we develop a graph convolutional encoder based on descendant neighbors, as well as a novel descendant prediction auxiliary task. Considering that a promoter can participate in the promotion of multiple products, we also incorporate the concept of hypergraphs to achieve a broader perspective while efficiently handling a large number of promotion networks. Regarding Challenge $2$, we maintain a promoter sub-table for each product, confining the training task to these sub-tables. It ensures the feasibility of the descendant prediction task while mitigating the sparsity issue. Additionally, an auxiliary task is introduced to alleviate high volatility by predicting whether promoters participate in promotion. Finally, after obtaining the predicted promotion structures and self-sales, the propagation scale can be successfully synthesized using descendant differentiable sampling and a matrix multiplication-based aggregation mechanism. The contributions of our work are summarized as follows:
\begin{itemize}
    \item We are the first to investigate the measurement of indirect propagation capabilities of promoters, introducing a novel metric called propagation scale and providing an open dataset to facilitate the following research on this problem.
    \item We propose a novel two-stage approach called DNTS for predicting propagation scale that leverages fundamental self-sales in conjunction with propagation structures. 
    % It includes a novel descendant-neighbor encoding technique and a descendant prediction task, and two strategies utilized to address the prevalent sparsity issues in industrial datasets.
    \item Extensive offline and online experiments on real industrial datasets demonstrate the superiority of our proposed method. At the same time, the experimental results reveal that the traditional utilization of GNNs in time series forecasting performs poorly in the affiliate marketing scenario, while DNTS can effectively use GNNs to boost performance (refer to Section~\ref{sec:5.6}).
\end{itemize}

\section{Related Work}
% \begin{yym revise}
% Previous work mainly focuses on the investigation of challenges and opportunities in affiliate marketing systems~\cite{af-1,af-2,af-3}. In this work, we focus on the problem of propagation scale forecasting in affiliate marketing in a computational way. Therefore, we review previous related methods from the perspective of technology.
Previous research mainly focuses on the challenges and opportunities faced by affiliate marketing from a business perspective~\cite{af-1,af-2,af-3}. In this work, we focus on the problem of predicting the promotion ability of promoters in a computational manner. 
Technically speaking, our work is related to time series forecasting, which is to predict future values based on historical observations. 
Given the critical role of promotion networks in our task, we selected graph-based time series forecasting as our core research focus. Earlier methods such as Graph WaveNet ~\cite{wavenet} and DCRNN~\cite{dcrnn} integrate graph diffusion convolutional layers into the temporal convolutional module to capture both temporal and spatial dependencies. STGCN~\cite{stgcn} and ST-MetaNet~\cite{stmetanet} employ GCN~\cite{gcn} or GAT~\cite{gat} to model spatial dependencies by aggregating information from adjacent time series. LSGCN~\cite{lsgcn} combines GAT and GCN in a gated manner to extract spatial adjacency more precisely. However, they heavily rely on pre-defined graph structure. To address this issue, MTGNN~\cite{mtgnn} introduces a graph learning layer that adaptively extracts adjacency relationships among variables. Follow-up work TPGNN~\cite{tpgnn} proposes a time-polynomial graph propagation to address the limitation of static variable dependencies in MTGNN. In addition, there are several Transformer-based models~\cite{tf-1,tf-2,tf-3}. They leverage Transformers to capture long-term dependencies to handle long-term time series forecasting.

Despite making considerable advancements, previous methods primarily cater to single-network scenarios. For example, in traffic prediction scenarios where time-series techniques are prevalently applied~\cite{lsnet,lsgcn,mtgnn,tpgnn,tpa-lstm}, sensors at various intersections in a given area have a fixed physical location,  corresponding to a single network. In contrast, we need to deal with $|I|*|T| $ distinct networks in affiliate marketing scenarios. Moreover, GNN-based methods~
\cite{stgcn,mtgnn,tpgnn,lsgcn}, directly perform convolutional operations on signals directly over graphs. Therefore, their effectiveness heavily relies on the assumption of structural smoothness of signals over the graph. Unfortunately, in the context of affiliate marketing, the signal distributions of propagation scale over the promotional network are not smooth with respect to the graph structure (refer to Section~\ref{sec:5.6}). This raises a concern about the validity of the existing utilization of GNNs in time series forecasting.

\section{Preliminaries}
We treat the behavior of one promoter retweeting a product link from another promoter as a promotion of that product. The resulting network linked by the retweets of a product is referred to as the promotion network. For the sake of convenience, we refer to products as items.

\begin{figure*}[ht]
\centering
\includegraphics[width=2.1\columnwidth]{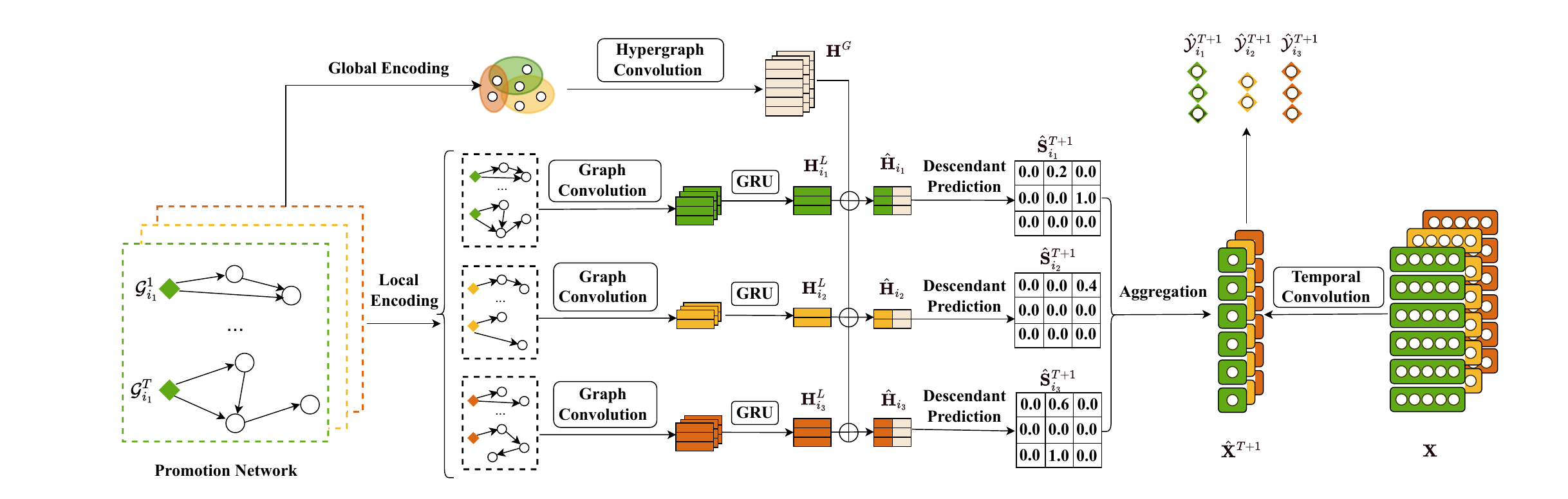}
\caption{The overall architecture of our proposed DNTS. Three items are displayed, marked in green, yellow, and red respectively.} 
\label{fig:model}
\end{figure*}

\begin{definition}
\textbf{Promotion Network.}  Let $\mathcal{I}$ denote the set of all items, and $\mathcal{M}$ denote the set of all promoters. Given an item $i$, its promotion network $\mathcal{G}_i = \{ \mathcal{M}_i, \mathcal{E}_i, \mathbf{x}_i \}$, where $\mathcal{M}_i$ is the set of promoters participating in the promotion of item $i$, $\mathcal{E}_i \subset \mathcal{M}_i \times \mathcal{M}_i $ is the set of directed edges, and $\mathbf{x}_i$ represents the signal vector, which stores self-sales of the promoters in $\mathcal{M}_i$. 
\end{definition}

\begin{definition}
\textbf{Promotion Network Snapshot.} Given that promoters typically follow a daily pattern in their promotion activities, we partition the promotion network into multiple snapshots by day. Specifically, the $t-th$ snapshot of $\mathcal{G}_i$ is defined as $\mathcal{G}_{i}^t = \{ \mathcal{M}_i^t, \mathcal{E}_i^t, \mathbf{x}_i^t \}$ that reflects the promotion status of $i$ on day $t$, where $\mathcal{M}_i^t$ is the set of promoters participating in promotion of $i$ on day $t$.
\end{definition}

\begin{definition}
\textbf{Propagation Scale.} To synthesize the propagation scale, we introduce external order data. Let $\mathcal{C}(o) = [m_s, \ldots, m_t]$ represent the attribution sequence of order $o$, where $m_s$ denotes the originator of the order and $m_t$ signifies the starting promoter of the promotion path. The set of orders associated with promoter $m$ is defined as $\mathcal{O}(m)$. Then, the propagation scale of promoter $m$ is represented as:
\begin{equation}
\label{label}
y({m}) = \sum_{o \in \mathcal{O}(m)} \phi(o),
\end{equation}
where $\phi(\cdot)$ is a function that returns the sales of an order, $y(\cdot)$ is a function that returns the propagation scale of promoters. We use a vector $\textbf{y}$ to store the propagation scale of promoters. 
\end{definition}

Referring to Figure~\ref{fig:case-2}(a), the attribution paths for two orders are marked in blue and red, with the corresponding sales of $2$, and $4$, respectively. Promoter $m_4$ is the generator of the blue order, resulting in a self-sales volume of $2$. The promoter $m_1$ is involved in the attribution sequence of both orders, thus its propagation scale is $4+2=6$. We can observe that the propagation scale of promoters results from the combination of self-sales from a subset of its descendants.

\begin{definition}
\textbf{Propagation Scale Forecasting.} Given a set of promotion network $\{ \mathcal{G}^{t_1}_{i}, \mathcal{G}^{t_2}_{i}, \ldots, \mathcal{G}^{T}_{i} \} $, the initial representations of promoters $\mathbf{H}^0 \in {{\mathbf{\mathbb{R}}}^{{|\mathcal{M}|} \times d_m}}$, and the initial representations of items $\mathbf{R}^0 \in {{\mathbf{\mathbb{R}}}^{{|\mathcal{I}|} \times d_i}}$, the problem is to predict a sequence of propagation scale for the future $\triangle t$ days $[\mathbf{y}_{i}^{T+1},\mathbf{y}_{i}^{T+2}, \ldots,\mathbf{y}_{i}^{T+\triangle t}]$.
\end{definition}

It is noteworthy that the order data is solely used during the preprocessing stage to construct our ground-truth. During the training phase, the model can access only the promotion networks and the self-sales of each promoter, but not the order information.

\section{Model}

\subsection{Temporal Convolution Module}
Inspired by MTGNN~\cite{mtgnn}, we introduce a temporal convolutional module with the Inception strategy~\cite{inception}, i.e., using filters with multiple sizes, which is able to discover temporal patterns with various ranges.

Specifically, this module consists of two inception layers: one layer is followed by a $\text{tanh}$ activation function as a filter, and another layer is followed by an $\text{sigmoid}$ activation function to gate the amount of information which can be transferred to the subsequent module, formally described as follows:
\begin{equation}
\mathbf{X}^{f} =  \text{Concat}(\text{tanh}(w_{1 \times k}^{f} * \mathbf{X}), k \in \mathcal{K}),
\end{equation}

\begin{equation}
\mathbf{X}^{g} = \text{Concat} (\text{sigmoid}(w_{1 \times k}^{g} * \mathbf{X}), k \in \mathcal{K}),
\end{equation}
where $\mathcal{K}$ is the set of different kernel sizes, $\mathbf{X} \in {{\mathbf{\mathbb{R}}}^{{|M|} \times T}}$ is obtained by concatenating $\mathbf{x}$ along the temporal dimension. Then, we can obtain the final forecasting results of self-sales as follows:
\begin{equation}
\label{eq:temporal-conv}
\hat{\mathbf{X}}=f_{\theta}(\text{Relu}(\mathbf{X}^{f} \odot \mathbf{X}^{g})),
\end{equation}
where $f(\cdot)$ represents a function composed of a stack of multiple convolution layers, $\theta$ denoting the parameters of the convolution kernels. In this way, the resulting temporal dimensionality is $\triangle t$.

\subsection{Spatial Convolution Module}

\subsubsection{Local Perspective}
% \begin{yymrevise}
After obtaining the fundamental self-sales forecasting results, we need to further predict the diffusion details to guide the synthesis of the propagation scale. However, reconstructing $|\mathcal{I}| \times |\mathcal{T}|$ promotion networks is impractical since the magnitude of $|\mathcal{I}|$ can easily reach the billion level in real industrial scenarios. Fortunately, it is practical to identify the potential descendants of each promoter. Therefore, we adopt the depth-first search (DFS) algorithm to obtain the descendant set of each promoter $m$ in a pre-processing manner, denoted as follows:
\begin{equation}
\label{eq:dfs}
\mathcal{D}(m)=\{d \mid \text{Path}(m \to d)=1, d \in \mathcal{M}\},
\end{equation}
% \begin{equation}
% \begin{split}
% \label{eq:dfs}
% \mathcal{D}(m)=&\{d \mid \text{Path}(m \to d)=1, d \in \mathcal{M}\}   \cup \\
% &\{d \mid \text{Path}(d \to m)=1, d \in \mathcal{M}\},
% \end{split}
% \end{equation}
where $\text{Path}(m \to d)=1$ means that there exists at least one reachable path from $m$ to $d$, in which case $d$ can be referred to as a descendant of $m$. At the beginning of each iteration, we randomly sample $\mathcal{D}$ to get the sampled set $\mathcal{D}^\prime$.

Then the graph convolution is performed using an inductive learning strategy, which can be formally described as follows:
\begin{equation}
\label{eq:graph-conv}
\begin{split}
&\textbf{h}_{d \to m} = \textbf{W}_2 \text{Concat}( \textbf{h}_d, \textbf{W}_1 \textbf{r}),\\
\textbf{h}_m^{\prime} = &\text{Att-Agg} (\left\{ \textbf{h}_{d \to m} \mid d \in \mathcal{D}^{\prime}(m) \right\})+\textbf{h}_{m},
\end{split}
\end{equation}
where $\textbf{h}_m^{\prime}$ represents the updated representation after convolution, $\text{Att-Agg}(\cdot)$ is an attention-based aggregation function. $\mathbf{W}_1 \in {{\mathbf{\mathbb{R}}}^{{d_r} \times d_m}}$ and $\mathbf{W}_2 \in {{\mathbf{\mathbb{R}}}^{{(2d_m)} \times d_m}}$ are parameter projection matrices. After obtaining the local representations of the promoters over $T$ days respectively, we adopt GRU to model the temporal dependencies as follows:
\begin{equation}
\textbf{h}^{Local}_m = \text{GRU} \left( {\textbf{h}^{\prime}_m}^{t_0},{\textbf{h}^{\prime}_m}^{t_1}, \ldots, {\textbf{h}^{\prime}_m}^{T} \right).
\label{eq:local}
\end{equation}
Note that the aforementioned operations are conducted within items. Therefore, the representation obtained from Eq.~(\ref{eq:local}) is specific to items and possesses a rather limited perspective. We denote $\textbf{h}_{m,i}^{Local}$ as the final local representation of promoter $m$ specific to item $i$.
% \end{yymrevise}

\subsubsection{Global Perspective}

Local encoding is employed to encode the promotional dynamics within an item. Since a promoter can participate in promoting multiple items, there are interactions between items. For instance, when promoter $m$ forwards shoes, it is highly likely that the next step will promote related items, such as socks. Consequently, in predicting $m$'s promotion network for socks, promotion information of shoes provides valuable cue.

To capture the interactions among different items, we introduce the concept of hypergraph. Specifically, we collect the promotion networks of all items on day $t$, obtaining $\mathcal{G}^t=\{\mathcal{G}_i^t \mid i \in \mathcal{I}\}$. Then we process $\mathcal{G}^t$ by treating promoters as nodes and their participation relationship in item promotion as hyperedges, resulting in an incidence matrix $\mathbf{B}^t \in \{0, 1\}^{|\mathcal{M}| \times |\mathcal{I}|}$, and $\mathbf{B}^t_{m,i} = 1$ indicates that $m \in \mathcal{M}^t_i$. Let $\tilde{e}_i = \{m|\mathbf{B}_{m,i}=1\}$ denote the hyperedge 
consisting of promoters participating in promoting $i$, and $\tilde{\textbf{r}}_{\tilde{e}_i}$ denote the representation of ${\tilde{e}_i}$. In order to distinguish from the local representation, we use $\tilde{\textbf{h}}$ to denote the global representation of promoters.

First, we learn the representation of the hyperedge $\tilde{{e}}_i$ by aggregating the representations of all the promoters associated with $\tilde{e}_i$, formally described as follow:
\begin{equation}
\label{eq:att}
   \tilde{\textbf{r}}^{\prime}_{\tilde{e}_i} =  \text{Concat}( \sigma(\sum_{m \in \tilde{e}_i} \alpha_{mi}  \tilde{\textbf{h}}_{m} ),\textbf{W}_1 \tilde{\textbf{r}}_{\tilde{e}_i}).
\end{equation}
The inputs to the first convolutional layer are $\tilde{\textbf{h}}_{m}=\textbf{h}^0_m$ and $\tilde{\textbf{r}}_{\tilde{e}_i}=\textbf{r}^0_i$. $\mathbf{W}_1$ is a parameter matrix, and $\alpha_{mi}$ is the aggregation weight obtained through the attention mechanism.

After obtaining representations of hyperedges, we train another aggregator to integrate all the hyperedges involved in $m$, described as follows:
\begin{equation}
\tilde{\textbf{h}}^{\prime}_{m} = \text{Relu}(\sum_{\tilde{e}_i \in \tilde{\mathcal{E}}(m)} \textbf{W}_2 \tilde{\textbf{r}}^{\prime}_{\tilde{e}_i}), 
\label{eq:hypergraph-conv}
\end{equation}
where $\tilde{\mathcal{E}}(m) = \{\tilde{e}_i|\mathbf{B}_{m,i}=1\}$, and $\mathbf{W}_2 \in {{\mathbf{\mathbb{R}}}^{{(d_m+d_r)} \times d_m}}$ is a parameter projection matrix. Similarly, GRU is utilized to model the temporal dependencies over $T$ days to obtain the global representation of promoter $m$:
\begin{equation}
\label{eq:globa}
\tilde{\textbf{h}}_m^{Global} = \text{GRU} \left( {\tilde{\textbf{h}}_m}^{\prime^{t_0}}, {\tilde{\textbf{h}}_m}^{\prime^{t_1}}, \ldots, {\tilde{\textbf{h}}_m}^{\prime^{T}} \right).
\end{equation}
Then, we can obtain the final representation of promoter $m$ under item $i$ as follows:
\begin{equation}
\label{eq:final-represention}
\hat{\textbf{h}}_{m,i}=\text{Concat}(\tilde{\textbf{h}}^{Global}_m,\textbf{h}^{Local}_{m,i}).
\end{equation}

\subsection{Decoding Phase}
We now possess the predicted self-sales ${\hat{\mathbf{X}}}$ and the structural representation of the promoters ${\hat{\mathbf{H}}}$. For clarity, we refer to $m$ in $\mathcal{D}(m)$ as the root promoter. Revisiting the example in Figure~\ref{fig:case-2}(a), the propagation scale of $m_1$ can be derived as follows:

\begin{align}
\label{eq:y_m}
\textbf{y}(m_1)=&s_{1,2}*\textbf{x}(m_2)+s_{1,3}*\textbf{x}(m_3)+s_{1, 4}*\textbf{x}(m_4)\nonumber\\
= &0.0*0+1.0*2+1.0*4=6
\end{align}

where $s_{m,d}$ represents what percentage of $d$'s self-sales are accounted for $m$'s propagation scale, referred to as the activation ratio of $m$ for $d$. 
The introduction of the activation ratio arises from the fact that a promoter's self-sales may originate from multiple channels, we need to ascertain how much of self-sales can be attributed to the root promoter.
According to Eq.~(\ref{eq:y_m}), to synthesize the propagation scale of $m_1$, it is essential to first identify the potential descendant set of $m_1$ (i.e, $m_2$, $m_3$, $m_4$) and subsequently determine the activation ratio within that set (i.e, $0.0$, $1.0$, $1.0$). 

To this end, a descendant prediction task is proposed to identify potential descendants. First, we use $\hat{\mathbf{H}}$ to compute the descendant probability scores. Then we employ the Gumbel-Softmax technique to perform differentiable sampling of potential descendants and then further predict the activation ratio based on the sampling results. The processes are formally expressed as follows:
\begin{equation}
\label{eq:softmax}
\mathbf{S}^{g}=\text{Gumbel-Softmax}(\text{sigmoid}({\hat{\mathbf{H}}}^T\mathbf{W}_3\hat{\mathbf{H}})),
\end{equation}
\begin{equation}
\mathbf{S}^{f}=\text{sigmoid}({\hat{\mathbf{H}}}^T\mathbf{W}_4\hat{\mathbf{H}}),
\end{equation}
where Gumbel-Softmax is a differentiable sampling technique, $\mathbf{S}^g \in {{\mathbf{\mathbb{R}}}^{{|\mathcal{M}|} \times |\mathcal{M}|}}$ stores the probabilistic score indicating whether a promoter is a descendant of the root promoter, which is utilized to gate the potential descendants. $\mathbf{S}^f \in {{\mathbf{\mathbb{R}}}^{{|\mathcal{M}|} \times |\mathcal{M}|}}$ records the predicted activation ratio. Thus, we can derive a coefficient matrix as follows: 
\begin{equation}
\hat{\mathbf{S}} = \mathbf{S}^g \odot \mathbf{S}^f, 
\label{eq:coeff}
\end{equation}
where the coefficients can be interpreted as weights used to aggregate the foundational self-sales to synthesize the propagation scale. A matrix multiplication-based aggregation mechanism is proposed to calculate propagation scale, which can be calculated as the following formula:
\begin{equation}
\label{eq:propagation-scale}
\hat{\mathbf{y}}^{T+t}={{\hat{\mathbf{S}}}^{T+t}} \cdot {\hat{\mathbf{X}}}{[:, t]}.
\end{equation}
% \end{yymrevise}

In this way, descendants with non-zero self-sales but an activation ratio of zero, as well as those with zero self-sales, will be excluded. Therefore it leaves only the descendants with both an activation ratio and self-sales greater than zero, which precisely represent the effective components for synthesizing the propagation scale, thus ensuring accurate predictions of the propagation scale.

% \begin{yymrevise}
\subsection{Model Loss}
After obtaining $\hat{\textbf{y}}$, we compute the loss for the main prediction task as the following formula:
\begin{equation}
\label{eq:loss_aa}
\mathcal{L}_{main} = \sum_{i \in \mathcal{I}} \sum_{t=1}^{\triangle t} \text{MSLE} (\mathbf{y}^{T+t}_i,\hat{\mathbf{y}}^{T+t}_i),
\end{equation}
where $\mathbf{y}^{T+t}_i$ is the ground-truth propagation scale. $\text{MSLE}(\cdot)$ is the mean squared logarithmic error, which is slightly different from the traditional metric of mean squared error (MSE) since the former can alleviate the impact of excessive propagation scale caused by promotional strategies.

\subsection{Addressing Unique Challenges in Affiliate Marketing}

\textbf{Data Sparsity}: There are billions of online items that need to be promoted, while each item has a finite audience of promoters. As a result, training a model based on all promoters leads to extremely sparse input data. Furthermore, the descendant prediction task based on all promoters is impractical due to an excessively large vocabulary size, which hinders softmax computation in Eq~\ref{eq:softmax}.

Fortunately, we observe that the promoter audience for an item is relatively stable. For example, for the clothing item in Fig~\ref{fig:case-2}(a), its audience promoters are often the same group of fashion bloggers over a period of time. Consequently, for each item, we collect all the related promoters who have participated in promoting it during a long time interval, to form a sub-table of item-specific promoters. Thus, during model training, we change the input format of the temporal convolution module changes from the dimensionality $b*t*|\mathcal{M}|*c$ into the dimensionality $b*t*|\mathcal{M}_i|*c$, where $b$ is the batch size, $c$ is channel number, $|\mathcal{M}|$ is the size of the all promoters, and $|\mathcal{M}_i|$ is the size of the promoter sub-table for item $i$. The local graph convolution module and descendant prediction task are carried out on the sub-table. In this way, we can effectively address the issue of data sparsity.

\textbf{High Volatility}: Items often experience a surge in demand over a few days due to promotional strategies such as increased commissions or lower prices. Therefore, the majority of promoters in $\mathcal{M}_i$ are drawn from those days of the burst period. It means that the scale of active promoters is significantly smaller than $|\mathcal{M}_i|$ most of the time. Therefore, $\hat{\mathbf{S}}$ should exhibit significant row sparsity, ideally. In contrast, $\hat{\mathbf{S}}$ derived from Eq.~(\ref{eq:coeff}) is generally dense, which poses challenges for decoding.

To address this, we establish a promoter activation prediction auxiliary task aimed at filtering out inactive promoters. In this way, we can avoid the ineffective computation of inactive promoters in advance, which improves computational efficiency. Furthermore, by forcing the coefficients corresponding to inactive promoters to be set to 0, we are able to obtain a sparser $\hat{\mathbf{S}}$, thus improving the decoding accuracy. Specifically, for those promoters whose active probability scores are below $\delta$, we directly update the corresponding rows in ${\hat{\mathbf{S}}}$ to zero, which can be expressed in the following formula:
\begin{equation}
{\hat{\mathbf{l}}}^{T+1}=\text{sigmoid}(\text{FC}({\hat{\mathbf{H}}})),
\end{equation}
\begin{equation}
{\hat{\mathbf{S}}}^{T+1} =({\hat{\mathbf{l}}}^{T+1} \odot \mathbb{I}({\hat{\mathbf{l}}}^{T+1} \geq \delta)) \odot {\hat{\mathbf{S}}}^{T+1},
\end{equation}
where $\hat{\mathbf{l}}$ is a vector that stores predicted activation scores, $\mathbb{I}$ is the indicator function used to filter out inactive promoters, $\text{FC}(\cdot)$ represents a full-connect layer. 

So far, we have configured two auxiliary tasks: descendant prediction and promoter activation prediction. To further improve the performance, an auxiliary task of self-sales prediction is introduced, which provides supervision information for the basic self-sales estimation. Thus, the loss of three auxiliary tasks can be described as follows: 
\begin{equation}
\label{eq:loss_aa}
\mathcal{L}_{aux} = \sum_{i \in \mathcal{I}} \sum_{t=1}^{\triangle t} \text{MSLE} (\mathbf{x}^{T+t}_i,\hat{\mathbf{x}}^{T+t}_i)+
\text{FL}(\mathbf{l}^{T+t}_i,\hat{\mathbf{l}}^{T+t}_i)+
\text{FL}({\mathbf{S}}^{T+t}_i, {\hat{\mathbf{S}}}^{T+t}_i),
\end{equation}
where $\mathbf{l}_i^{T+t}$ represents the true label indicating whether the promoter is activated by item $i$ on the day $T+t$, with values being $0$ or $1$. ${\hat{\mathbf{S}}}^{T+t}$ denotes the true descendant relationship matrix of $\mathcal{G}^{T+t}_i$, whose values are binary. $\text{FL}(\cdot)$ is the focal loss that can alleviate the label imbalance issue. Finally, combining the loss terms of both the main task and the auxiliary tasks, the overall loss can be described as follows:
\begin{equation}
\label{eq:loss_all}
\mathcal{L} = \mathcal{L}_{main}+\mathcal{L}_{aux}.
\end{equation}

\section{Experiments}

\subsection{Datasets}
We select $1000$ of the most popular products from the Alimama platform and collect their sales data and promotion logs from $2024/07/14$ to $2024/08/13$, denoted as $\textit{30-days}$. To explore the performance of the model over different time spans, we additionally derive two datasets, marked as \textit{7-days} (from $2024/07/14$ to $2024/07/20$), \textit{15-days} (from $2024/07/14$ to $2024/07/28$). Unlike previous methods, our objective is to predict the propagation scale of a specific item. Therefore, we divide the data into training, validation, and test sets along the dimension of the item, with proportions of $60\%$, $10\%$, and $30\%$, respectively. Specifically, the \textit{7-day} dataset encompasses $87350$ promoters and $1568214$ edges in promotion networks; the $\textit{15-day}$ dataset includes $102776$ promoters and $3365140$ edges, and the \textit{30-day} dataset comprises $104557$ promoters and $5928263$ edges.

\subsection{Baselines}
Since the promotion network is a particularly important feature for our task, we select several time-series baselines that are able to handle graph structures: DCRNN~\cite{dcrnn}, STGCN~\cite{stgcn}, LSGCN~\cite{lsgcn}, MTGNN\cite{mtgnn} and TPGNN~\cite{tpgnn}. However, they cannot handle multiple dynamic networks. To address this issue, we integrate promotion networks over $T$ days for each item. For instance, for item $i$, the collection of promotion networks over $T$ days $\{ \mathcal{G}^{t_1}_{i}, \mathcal{G}^{t_2}_{i}, \ldots, \mathcal{G}^{T}_{i} \}$ is merged, and the corresponding promoter set is $\bigcup_{t=1}^{T} \mathcal{M}^t_i$. For promoters that do not participate in promotion on day $t$, we set the corresponding values in $\mathbf{x}^t_i$ to 0. In this way, the dynamic network is transformed into a static one, allowing baselines to function effectively.

\subsection{Experimental Settings}
 For the three datasets with varying temporal spans, we consistently reserved the final day's data for testing while utilizing the preceding days for model training. And MSLE and MAPE are adopted as the evaluation metric. A smaller MSLE indicates better performance, and the same applies to MAPE. We summarize the subsequent experiments into four modes, described as follows:
\begin{itemize}
    \item $P \to P$: Both input and output are propagation scale. All GNN-based baselines are categorized under this mode, which performs convolution operations on the signals based on static spatial dependencies.
    \item $S \to S$: Both input and output are self-sale.
    \item $S \dashrightarrow P$: Input is self-sale, and output is propagation scale, employing a two-stage scheme, i.e., the DNTS we proposed.
\end{itemize}

\subsection{Main Results }

The experimental results are shown in Table~\ref{tab:main-result}. We can observe that DNTS shows significant performance improvements over all baselines, with an average enhancement of $12\%$ in MSLE and $6\%$ in MAPE compared to the best baseline TPGNN, highlighting its superiority. And all methods achieve optimal performance with the $\textit{30-days}$ dataset, outperforming both $\textit{15-days}$ and $\textit{7-days}$, which is attributed to the increase in the number of historical observations available in $\textit{30-days}$.

ARIMA performed the worst due to its limitations as a statistical model in handling complex non-linear time series data. LSGCN outperforms STGCN in long-term predictions by employing a simpler network structure with fewer layers, reducing cumulative errors from iterative predictions. MTGNN outperforms both LSGCN and STGCN, potentially attributed to its graph learning module that automatically extracts relations among variables and advanced dilated inception convolution. TPGNN enhances performance further by introducing polynomial graph convolution to address the static variable dependency issue in MTGNN.

\begin{figure}[ht]
\centering
\includegraphics[width=1.0\columnwidth]{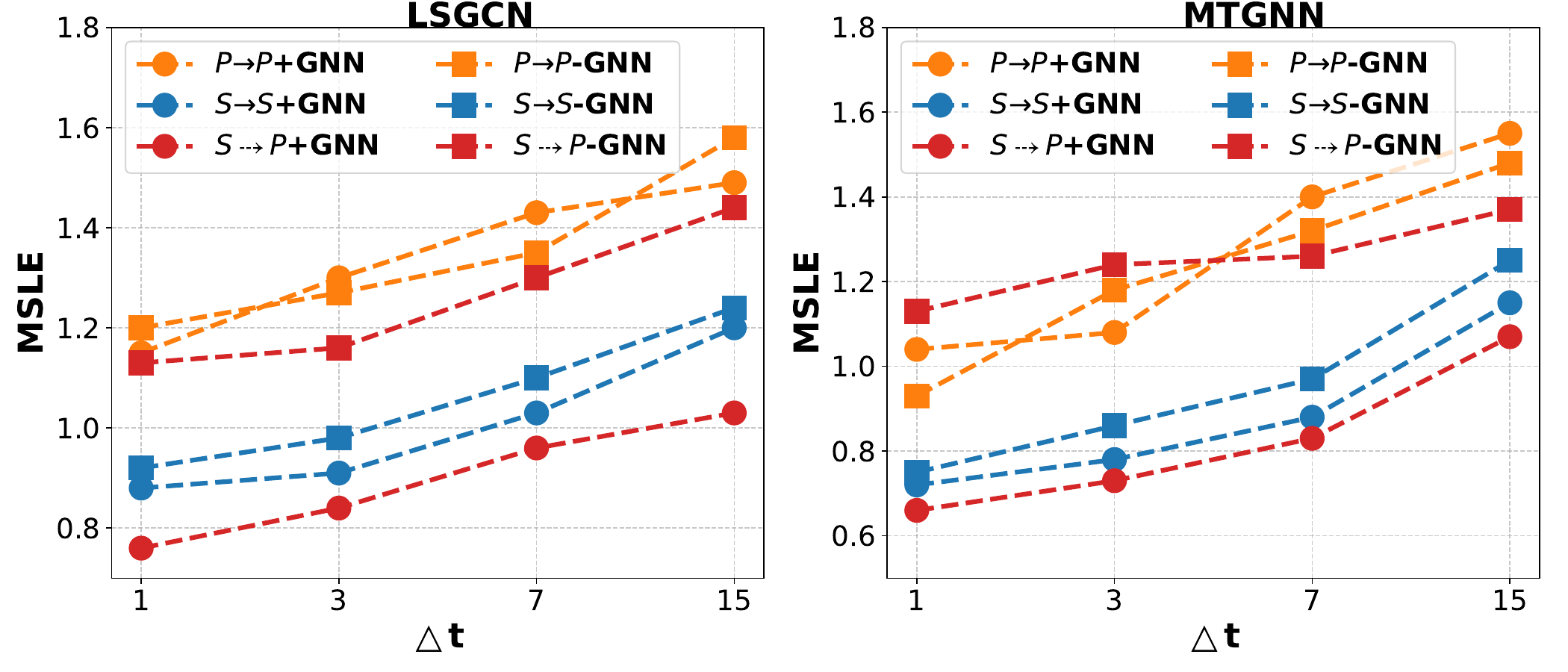}
\caption{"+GCN" denotes the original model architecture, while "-GCN" signifies its variant without the graph convolutional module. The red dotted lines in both subfigures represent our proposed DNTS, the other dotted lines are baselines. All experiments are conducted on \textit{30-days}.} 
\label{fig:two-step}
\end{figure}
\begin{table}

	\centering
	\caption{The overall performance comparison.}
        \label{tab:main-result}
            {\renewcommand{\arraystretch}{0.9}
	{
		\begin{tabular}{c l l c c c}
			\toprule
			Task & Methods & Metrics  &\textit{7-days} & \textit{15-days} & \textit{30-days}
			\\
			\midrule

   \multirow{13}{*}{$P \to  P$} & \multirow{2}{*}{ARIMA} & MSLE & 1.48 & 1.37 & 1.36\\
	   &&MAPE$(\%)$&110.36&108.61&108.02\\	
 \cline{2-6} 
    &\multirow{2}{*}{DCRNN}  & MSLE & 1.33 & 1.30 & 1.22   \\
   &&MAPE$(\%)$&92.73&86.39&86.31\\
        \cline{2-6}   
&\multirow{2}{*}{STGCN} & MSLE  & 1.17  & 1.12 & 1.10 \\

&&MAPE$(\%)$&77.90&78.44&72.50\\
        \cline{2-6}   
&\multirow{2}{*}{LSGCN} & MSLE & 1.26 & 1.06 & 1.03 \\

	   &&MAPE$(\%)$&80.63&78.41&73.66\\
        \cline{2-6}   
    &\multirow{2}{*}{MTGNN} & MSLE & 1.10 & 1.02  & 1.02 \\
    &&MAPE$(\%)$&71.27&70.87&69.47\\
        \cline{2-6}   
    &\multirow{2}{*}{TPGNN}  & MSLE & 1.05 & 1.02 & 0.97\\
	   &&MAPE$(\%)$&70.32&70.15&69.14\\ 	
        \hline 
\multirow{2}{*}{$S \dashrightarrow P$} &\multirow{2}{*}{\textbf{DNTS}}  & MSLE & \textbf{0.92} & \textbf{0.90} & \textbf{0.86}   \\
&& MAPE$(\%)$& \textbf{68.70}  & \textbf{66.37} &  \textbf{66.27}   \\
			\bottomrule
		\end{tabular}}
	}
\end{table}

\subsection{Effect of Graph Structure on Performance}
\label{sec:5.6}
In this study, we aim to explore whether the integration of structural information modeled by GNNs universally enhances the forecasting capability for propagation scale or self-sale. To this end, we conduct three task groups: $P \to P$, $S \to S$, and $S \dashrightarrow P$, each treating the network structural information as a control variable. 
 Notably, the latter utilizes graph information differently than the first two groups: $P \to P$  and $S \to S$ are based on a static graph for signal convolution while $S \dashrightarrow P$
 independently models the dynamic graph structure.

The results, depicted in Figure~\ref{fig:two-step}, present an intriguing phenomenon: while the incorporation of structural information enhances the performance of self-sales prediction, its impact on the prediction of propagation scale is not consistently positive. This aligns with our previous analysis that the effectiveness of GNNs relies on the smoothness assumption -  neighboring nodes should have similar signal intensity, which is not fulfilled by the distribution of propagation scale over the promotion network. 

Moreover, GNNs only provides a slight improvement in predicting self-sales. Hence, we can conclude that the GNN usage of convolving the signal directly on the graph is of limited effect in affiliate marketing scenarios. In contrast, Figure~\ref{fig:two-step} shows that $S \dashrightarrow P + GNN$ has a significant improvement compared to $S \dashrightarrow P - GNN$, which implies the two-stage scheme in our proposed DNTS allows GNNs to realize the greater potential.

\begin{table}[htbp]
    \centering
    \caption{Results from the online A/B test. The data in the table is the aggregated totals over a one-month period.} 
    \label{tab:online}
    \begin{tabular}{l  c  c  c} 
        \hline 
        Metric & Baseline Bucket & Experiment Bucker & Improv$\uparrow$ \\
        \hline
        GMV    & $7.74 \times 10^7$ & $8.46 \times 10^7$ & $+9.29\%$ \\
        \hline 
        Sales  & $1.51 \times 10^6$ & $1.60 \times 10^6$  &  $+5.89\%$ \\
        \hline
    \end{tabular}
\end{table}

\subsection{Online Serving \texorpdfstring{$\&$}{\&} A/B Testing.}
We deployed DNTS on the Alimama recommendation platform to optimize the daily recall stage. Specifically, the recall score $r(p,i)$ of item $i$ for promoter $p$  is updated as $w(p,i)*r(p,i)$, where $w(p,i)=(1+sigmoid(y(p,i))$, and $y(p,i)$ is the promotion scale predicted by DNTS. This configuration was implemented in the experiment bucket. For comparison, we employed a strong baseline MTGNN~\cite{mtgnn}, applying similar processing in the baseline bucket. To mitigate potential population bias, we conducted a one-month evaluation, with the latter half involving a swap of user groups between the two buckets. We measured two key metrics, GMV and sales, as summarized in Fig~\ref{tab:online}. We can find that since DNTS can more accurately predict the promoters’ propagation scale for specific items, it improves the recall quality of the experimental barrel, resulting in substantial gains in both GMV and sales.

\section{Conclusion}
This paper addresses for the first time the evaluation and prediction of promoter capabilities in affiliate marketing, introducing a novel evaluation metric termed propagation scale and proposing a time series prediction model DNTS. Unlike conventional one-step prediction, DNTS innovatively adopts a two-stage scheme: initially forecasting self-sales and promotion networks separately, followed by the synthesis of the propagation scale. Extensive offline and online experimentation on industrial datasets 
validates the superiority of the proposed DNTS.

\section*{Acknowledgments}
\begingroup 
\setlength{\emergencystretch}{3em}
This work was supported in part by the National Natural Science Foundation of China under Grants 62425605, 62133012, and 62303366, in part by the Key Research and Development Program of Shaanxi under Grants 2022ZDLGY01-10, 2024CY2-GJHX-15, and 2025CY-YBXM-041, and in part by the Xidian University Specially Funded Project for Interdisciplinary Exploration under Grant TZJHF202506.
\endgroup
\section*{Generative AI Disclosure Statement }
All the content of this paper, including data collection, code implementation, and writing, did not use any generative AI tools.

\bibliographystyle{ACM-Reference-Format}
\balance

\bibliography{refs.bib}

\end{document}